\documentclass[twocolumn,pre,floats,aps,amsmath,amssymb,nofootinbib]{revtex4-2}
\usepackage{graphicx}
\usepackage{bm}
\usepackage{physics}
\usepackage[labelformat=empty]{caption}

\begin{document}

\title{Wigner’s Friend Depends on 
Self-Contradictory Quantum Amplification}
\author{Andrew Knight}
\affiliation{aknight@alum.mit.edu}
\date{\today}

\begin{abstract}
In a recent paper, \.Zukowski and Markiewicz showed that Wigner's Friend (and, by extension, Schr\"odinger’s Cat) can be eliminated as physical possibilities on purely logical grounds.  I validate this result and demonstrate the source of the contradiction in a simple experiment in which a scientist S attempts to measure the position of object $\ket{O} = \ket{A}_S + \ket{B}_S$ by using measuring device M chosen so that $\ket{A}_M \approx \ket{A}_S$ and $\ket{B}_M \approx \ket{B}_S$.  I assume that the measurement occurs by quantum amplification without collapse, in which M can entangle with O in a way that remains reversible by S for some nonzero time period.  This assumption implies that during this ``reversible" time period, $\ket{A}_M \neq \ket{A}_S$ and $\ket{B}_M \neq \ket{B}_S$ – i.e., the macroscopic pointer state to which M evolves is uncorrelated to the position of O relative to S.  When the scientist finally observes the measuring device, its macroscopic pointer state is uncorrelated to the object in position $\ket{A}_S$ or $\ket{B}_S$, rendering the notion of ``reversible measurement" a logical contradiction.
\end{abstract}

\maketitle

Schr\"odinger’s Cat \cite{Schrodinger} just won’t die.  Its conscious counterpart, Wigner’s Friend \cite{Wigner}, has also made a recent resurgence in the physics academy \cite{Bong,Brukner,Frauchiger,Proietti,Relano}.  Ordinarily, physicists treat such issues as empirical questions, relegating armchair philosophy to the humanities.  That’s why, as a legal scholar, I was surprised and delighted that \textit{Physical Review Letters} published \cite{Zukowski}, in which \.Zukowski and Markiewicz argue that the Wigner's Friend ("WF") paradox can be eliminated on \textit{a priori} logical grounds because it depends on the “inherently self-contradictory notion” that WF measures outcomes.  In essence, the authors show that a reversible measurement is a logical contradiction.  If they are correct, their work may have implications not only for the physical viability of any macroscopic quantum superposition “beyond the Heisenberg cut,” including Schr\"odinger’s Cat (``SC"), but also for physical limits to quantum computing and whether decoherence \cite{Joos} causes objective wave function collapse.  

Logic-based algorithms do occasionanlly sneak into debates about the nature of the physical world.  For example, \cite{Hagar} argues that the case for scalable quantum computing rests on a logical contradiction, \cite{Barros,Roselli} argue that the consciousness-causes-collapse hypothesis is logically immune from empirical falsification, while \cite{Knight} argues that a logical contradiction infests literature on the characterization of quantum superpositions.  Can a purely logical argument tell us something about the physical possibility of SC/WF?  No one has ever created SC/WF, despite the confusing use of “cat state” to describe certain quantum superpositions \cite{Yin}.  The assertion that SC/WF is a physical possibility is itself a logical inference arising from an assumption of quantum amplification without collapse.  If that assumption can be shown to be self-contradictory on logical grounds, then indeed a logical argument can tell us something about SC/WF.  

Many of the big problems in foundational physics, including the measurement problem \cite{Gao,Maudlin} and SC/WF, arise from a universality assumption (“U”) that the quantum mechanical wave function always evolves linearly or unitarily or reversibly, an assumption that seems to be rarely challenged in the literature.\footnote{The authors of \cite{D'Ariano,GRW,Hobson,Kastner,Penrose} are among these maverick challengers.}  According to the mathematics of quantum mechanics, when an object in a quantum superposition interacts with another system, such as a measuring device, the two “entangle” or “correlate” as if the system inherits the superposition.  Unless the superposition objectively collapses (contrary to U), it gets inherited by other systems with which the system interacts, leading to an amplification of the original superposition.

Consider a cat in an isolated box.  The box contains a radioisotope (which is in a quantum superposition over states “decayed” and “undecayed”) and a device that releases a poison if the radioisotope decays.  U implies that the device inherits the radioisotope’s superposition, then the cat inherits the device’s superposition, leading to that confusing hypothetical SC state in which the cat seems to be in a superposition over states “dead” and “alive.”  Quantum amplification without collapse, which I’ll describe in more detail below, is crucial for the possibility of SC.

A scientist S (initially in state $\ket{S}$) wants to measure the position of a tiny object O, which is in a superposition of macroscopically distinct semiclassical position eigenstates corresponding to locations A and B separated by some distance $d$.  Neglecting normalization constants, $\ket{O} = \ket{A} + \ket{B}$.  To measure it, he uses a measuring device M (initially in state $\ket{M}$) configured so that a measurement of the position of O will correlate M and O such that M will then evolve over time to a corresponding macroscopic pointer state, denoted $\ket{M_A}$ or $\ket{M_B}$.  Device M in state $\ket{M_A}$, for example, indicates “A” such as with an arrow-shaped indicator pointing at the letter “A.”  In other words, M is configured so that if it measures O at location A then M will, through a semideterministic causal chain that amplifies the measurement, evolve to some state that is obviously different to scientist S than the state to which it would evolve had it measured O at location B.  

Let’s say that the experiment is set up at time $t_0$; then at time $t_1$ M “measures” O via some initial correlation event, after which M then evolves in some nonzero time $\Delta t$ to a correlated macroscopic pointer state; then at $t_2$ S reads the device’s pointer.  Time $\Delta t$ is small, so in the following equation I’ll ignore it.  The scientist in state $\ket{S_A}$, for example, (believes he) has measured the position of O at position A by nature of having observed the device in state $\ket{M_A}$.  The standard narrative in quantum mechanics, following from the assumption of U, is this von Neumann \cite{Neumann} amplification chain:
\begin{eqnarray}
&t_0:& \ket{O}\ket{M}\ket{S} \nonumber\\
&=& \Bigl(\ket{A}+\ket{B}\Bigr) \ket{M}\ket{S} \nonumber\\
&\stackrel{t_1}{\longrightarrow}& \Bigl(\ket{A}\ket{M_A}+\ket{B}\ket{M_B}\Bigr) \ket{S} \nonumber\\
&\stackrel{t_2}{\longrightarrow}& \ket{A}\ket{M_A}\ket{S_A}+\ket{B}\ket{M_B}\ket{S_B} 
\end{eqnarray}

According to Eq.\ 1, at time $t_1$, object O and device M are entangled but scientist S remains separable, while at time $t_2$, all three are entangled.  What this means physically is that between times $t_1$ and $t_2$, S can – at least “in principle” – reverse the correlation event between O and M by performing a properly designed interference experiment.  The state at $t_1$ is the elusive “cat state” in which M seems to be in a superposition over state $\ket{M_A}$, in which it macroscopically indicates “A,” and state $\ket{M_B}$, in which it indicates “B.”  The possibility of producing SC depends on the assumption that a quantum object can entangle with a cat in a way that remains physically reversible by an external observer for some nonzero time.  SC cannot exist, even in principle, unless Eq.\ 1 is correct.  Is it?  Eq.\ 1 implies:
\begin{itemize}
\item There is an event between O and M at $t_1$, and 
\item S can physically reverse this event at any time before $t_2$.
\end{itemize}

However, if S can reverse the event, then in what sense, by whose observation, and by whose clock does the event occur?  

At $t_1$, can S correctly claim that the event occurred?  No.  At $t_1$, S is uncorrelated to M and O and has absolutely no evidence that they interacted.  He \textit{can’t} have any evidence, because evidence would correlate them.   He can’t say, at $t_1$, that he \textit{will} have evidence at $t_2$, because he can reverse the event before $t_2$.  In the reversible period between $t_1$ and $t_2$, there is not, from the perspective of S, an event between O and M.

At $t_1$, does the event occur from the perspective of M?  Again, no.  Ref.\ \cite{Zukowski} shows that “Outcomes of premeasurements… are an internally inconsistent notion (and therefore are not ‘facts’).”  And there is another reason.  What does time $t_1$ even mean?  By what standard?  Well, by a clock, of course!  A clock is just an object that physically ticks off time – a wristwatch, a cellular clock, a vibrating piece of quartz, etc.  If you were to take an isolated system containing, say, an old cat and a grandfather clock, and perfectly reversed the entire system\footnote{And I mean \textit{perfectly}.  The trajectory of every atom would have to be reversed with perfect precision.  It’s not possible, because even the largest objects in the universe (black holes) are swayed by unpredictable quantum chaos \cite{Boekholt}, but it helps with the thought experiment.} until the cat was young, then, from the cat’s perspective, was the cat ever old?  No.\footnote{See interesting discussions in \cite{Aaronson,Aaronson2,Maccone}.}  By perfectly reversing the system, every piece of evidence that might indicate the passage of time is retroactively erased.  The clock, of course, indicates no passage of time, but neither does the cat’s body or brain.  If you perfectly reverse an isolated system, you perfectly reverse its time, too.\footnote{Deutsch \cite{Deutsch} argues that WF can be produced without perfectly reversing time by having WF write a letter to Wigner that is uncorrelated to the outcome that he ``observed."  Salom \cite{Salom} correctly points out Deutsch's error.}

On this basis, Eq.\ 1 is false because it implies two contradictory statements.  How can there be an event at $t_1$ if S can, before $t_2$, physically reverse it as if it never happened?  Ref.\ \cite{Zukowski} is correct: a reversible measurement is a contradiction.  The von Neumann amplification chain of Eq.\ 1, on which the in-principle possibility of SC rests, is internally inconsistent and thus false.  

The problem with Eq.\ 1, as I will explain below with a simple example, is that the appearance of the nonzero time period ($t_2 - t_1$) is an illusion caused by failure to keep track of what the letters “A” and “B” actually refer to in each of the terms.  In the above example, $\ket{A}$ is an eigenstate of the position operator corresponding to localization at position A.  But Galileo and Einstein might point out that “position A” only has meaning relative to other objects.  For scientist S using measuring device M, is position A relative to S or to M?  Is the object O in Eq.\ 1 in state $\ket{A}_S+\ket{B}_S$ or $\ket{A}_M+\ket{B}_M$?  Ordinarily, it would not matter.  After all, when would the scientist and his measuring device disagree on the location of position A?  

Imagine two measuring devices that are in a superposition relative to each other \cite{Aharonov,Giacomini,Loveridge,Zych}.  The first one might see position A as a sharp point while the second might see it ``blurred."  If the blurred view still allows the second measuring device to distinguish position A from position B, then there’s no problem.  But if the width of the “blurriness” exceeds the distance $d$ between A and B, then “position A” as measured by the first device will not correlate to “position B” as measured by the second.  So if the scientist and his measuring device were spatially “blurry” relative to each other, they might disagree about “position A,” in which case the state of O would have to be written as either $\ket{A}_S+\ket{B}_S$ or $\ket{A}_M+\ket{B}_M$.  

In reality, it’s hard to see how this could happen.  Quantum dispersion is a natural process that makes all objects spatially fuzzy relative to each other over time.  A consequence of quantum uncertainty, loosely defined as $\Delta x(m \Delta v) \ge \hbar /2$, dispersion is inversely related to mass as well as tempered by decohering interactions with particles zipping around the universe \cite{Joos}.  The larger an object is, the more slowly it disperses and the more opportunity it has to be localized by another object.\footnote{Tegmark \cite{Tegmark}, for example, calculates that the coherence length of a $10 \mu m$ dust particle, due to decoherence by cosmic microwave background radiation ubiquitous throughout the universe, is $10nm$, a full three orders of magnitude smaller than the dust particle itself.  Physically, this means that such a dust particle could not, in the actual universe, disperse adequately to subject it to a double-slit interference experiment.}  So while the scientist might in principle be delocalized from the measuring device by some miniscule amount, that amount is much, much smaller than the distance $d$ distinguishing positions A and B.  As long as position A relative to the scientist is approximately the same as position A relative to the device, then $\ket{O} = \ket{A}_S+\ket{B}_S \approx \ket{A}_M+\ket{B}_M$.\footnote{$\ket{A}_S \approx \ket{A}_M$, for instance, does not mean that location A is the same distance from S as from M.  Rather, it means that location A relative to S is well correlated to location A relative to M; there is zero (or inconsequential) quantum fuzziness between the locations; S and M ``see" see point A as essentially the same sharply defined location.}

Let’s look again at the implications of Eq.\ 1 and how its internal contradiction arises in a simple example in which the distinction between states $\ket{A}_S$ and $\ket{A}_M$, for example, is respected.  In Fig.\ 1a, object O is shown at time $t_0$ in a superposition of position eigenstates $\ket{A}_S$ and $\ket{B}_S$ corresponding to locations at $A_S$ and $B_S$ relative to scientist S, where O is shown crosshatched to represent its superposition relative to S.  Measuring device M has slots (a) and (b) and is configured so that detection of object O in slot (a) will, due to a semi-deterministic causal amplification chain, cause it to evolve over nonzero time $\Delta t$ to a macroscopic pointer state in which a large arrow indicator points to the letter “A,” and vice versa for detection of object O in slot (b).  S intends to measure the object’s location relative to \textit{him} and wants the device’s indicator to correlate to that measurement.  Therefore, because the device’s detection of O in slot (a) actually corresponds to measurement of O at location $A_M$ relative to M (and vice versa for slot (b)), S configures M at time $t_0$ so that $A_M \approx A_S$ and $B_M \approx B_S$.  The experiment is designed so that the initial correlation event between O and M occurs at time $t_1$, M evolves to its macroscopic pointer state by time $t_1+\Delta t$, and S reads the device’s pointer at $t_2$.

Ignoring Eq.\ 1 for the moment, Figs.\ 1b and 1c show how the scientist might want the system to evolve.  In Fig.\ 1b (at time $t_1$), as long as $A_M \approx A_S$ and $B_M \approx B_S$, the device’s detection of O in slot (a) correlates to the object’s location at $A_S$.\footnote{No collapse or reduction of the wave function is assumed in this example or anywhere else in this paper.  Fig.\ 1b simply shows the correlation between detection in slot (a) and position $A_S$; there would also be a correlation (not shown) between detection in slot (b) and position $B_S$.}  Then, in Fig.\ 1c (at time $t_1+\Delta t$), M has evolved so that the indicator now points to letter “A,” correlated to the device’s detection of O in slot (a).  Then, when S looks at the indicator at time $t_2$, he will observe the indicator pointing at “A” if O was localized at $A_S$ and “B” if it was localized at $B_S$, which was exactly his intention in using M to measure the object’s position.
\begin{figure}[ht]
\centering
\includegraphics[width=2.0 in]{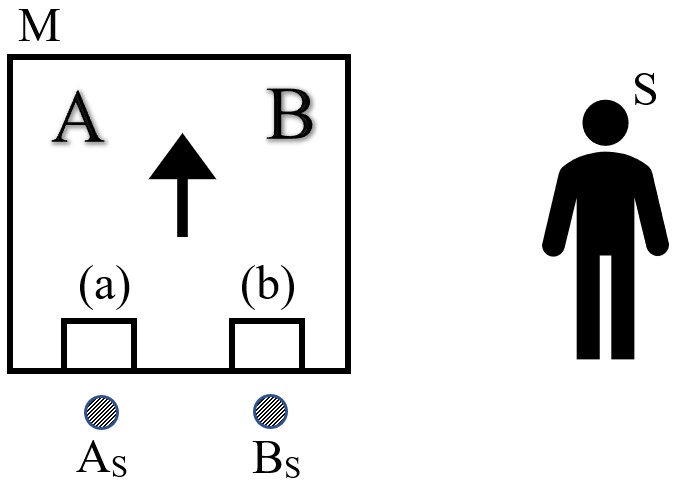}
\caption{Fig.\ 1a. An object is shown at time $t_0$ in a superposition of position eigenstates $\ket{A}_S$ and $\ket{B}_S$ corresponding to macroscopically distinct positions $A_S$ and $B_S$ relative to scientist S, who uses measuring device M to measure the object’s position.}
\end{figure}
\begin{figure}[ht]
\centering
\includegraphics[width=2.0 in]{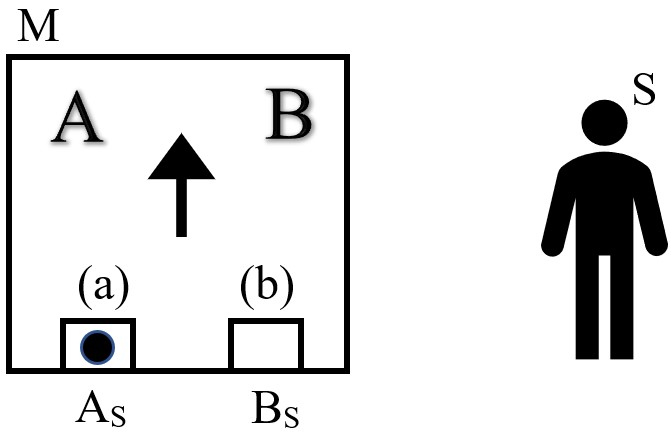}
\caption{Fig.\ 1b. The object is measured at time $t_1$ by M so that detection in slot (a), which corresponds to localization at position $A_M$ relative to M, also correlates to position $A_S$ relative to S.  The situation of detection in slot (b), corresponding to localization at position $B_M$, is not shown.}
\end{figure}
\begin{figure}[ht]
\centering
\includegraphics[width=2.0 in]{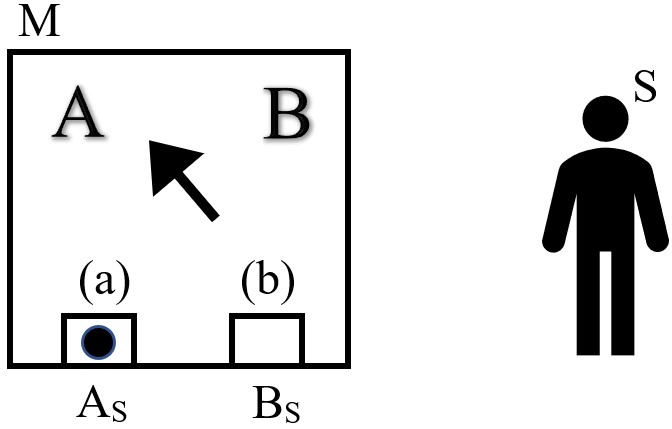}
\caption{Fig.\ 1c. By time $t_1+\Delta t$, M evolves to a macroscopic pointer state in which an indicator points to letter “A,” read by S at time $t_2$.}
\end{figure}

In Fig.\ 1b, O is indeed localized relative to M, and since $A_M \approx A_S$ and $B_M \approx B_S$, it is also localized relative to S.  The scientist does not know, of course, whether object O was detected in slot (a) or (b), and Fig.\ 1b only shows the first possibility, but it \textit{is} in slot (a) or (b), with slot (a) correlated to $A_S$ and slot (b) correlated to $B_S$ (both of which are also localized relative to S).   Because O is localized relative to S at $t_1$, S and O are correlated, in which case S cannot do an interference experiment to show O or M in superposition, which conflicts with the requirements of Eq.\ 1.

Let’s imagine a slight variation to Fig.\ 1b, shown in Fig.\ 2, in which S has misplaced M so that its measurement of O at position $A_M$ (by nature of detecting the object in slot (a)) correlates to position $B_S$ relative to the scientist.  When the scientist later sees the indicator point to the letter “A,” the result is certain to be wrong.  However, certainty implies correlation; if the scientist ever realizes that he misplaced the device, he will know that the measurement outcome was actually B.
\begin{figure}[ht]
\centering
\includegraphics[width=2.0 in]{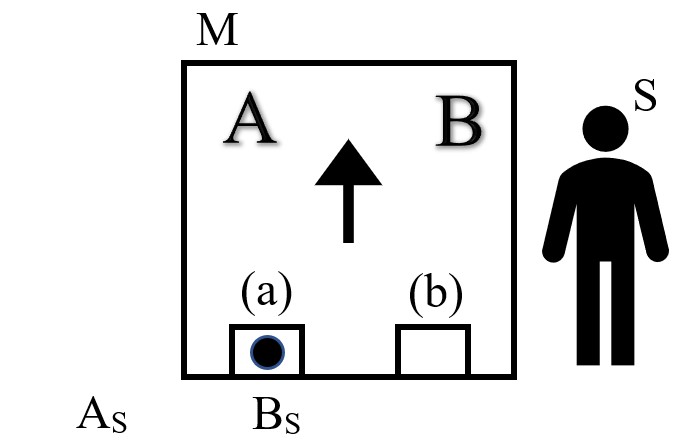}
\caption{Fig.\ 2.  In a variation on Fig.\ 1b, M has been misplaced so that detection in slot (a), which corresponds to localization at position $A_M$ relative to M, correlates to position $B_S$ relative to S.  The situation of detection in slot (b), corresponding to localization at position $B_M$, is not shown.}
\end{figure}

Now, suppose we demand, consistent with Eq.\ 1, that at time $t_1$, scientist S can, with an appropriately designed interference experiment, demonstrate that O and M are in a superposition relative to S.  That requirement implies that the object’s location, as measured by M via the correlation event at $t_1$, does not correlate to the object’s location relative to S.  The device’s detection at $t_1$ of O in, for example, slot (a), which corresponds to its measurement of the object at $A_M$, cannot correlate to the object’s location at $A_S$.  Thus, to ensure that O remains unlocalized relative to S when the correlation event at $t_1$ localizes O relative to M, that location which M measures as $A_M$ (by detection of object O in slot (a)) cannot correlate to location $A_S$.  By combining Figs.\ 1b and 2, this situation is shown in Fig.\ 3a in which both O and M are shown in a superposition of position eigenstates relative to S.  
\begin{figure}[ht]
\centering
\includegraphics[width=2.0 in]{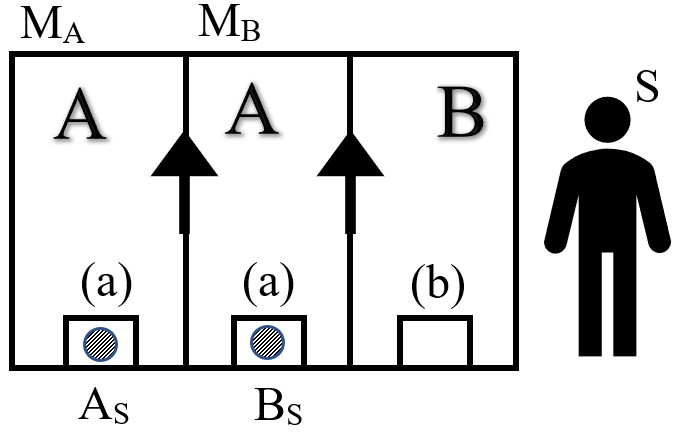}
\caption{Fig.\ 3a.  Combining Figs.\ 1b and 2, both O and M are shown at time $t_1$ in a superposition of position eigenstates relative to S.  Detection in slot (a) by M in position $M_A$ correlates to O in position $A_S$ while detection in slot (a) by M in position $M_B$ correlates to O in position $B_S$.  The situation of detection in slot (b) is not shown.}
\end{figure}

In Fig.\ 3a, the object’s crosshatching, like in Fig.\ 1a, represents its superposition over locations $A_S$ and $B_S$.  Analogously (but without crosshatching), device M localized at $M_A$ is shown superimposed with device M localized at $M_B$.  Importantly, $M_A$ is the position of M, relative to S, that would measure the position of object O at $A_M$ (by detecting it in slot (a)) as position $A_S$, while $M_B$ is the position of M that would measure the position of object O at $A_M$ (by detecting it in slot (a)) as position $B_S$.  Because O remains uncorrelated to S at $t_1$ (as demanded by Eq.\ 1), the device M that detects O in slot (a) must also be uncorrelated to S at $t_1$.  (The same is true for device M that detects O in slot (b), which is not shown.)  What is demonstrated in Fig.\ 3a is that the correlation event at $t_1$ between M and O that localizes O relative to M requires that M is not correlated to S, thus allowing S to demonstrate M in a superposition, as required by Eq.\ 1.  Then, in Fig.\ 3b, at time $t_1+\Delta t$, M has evolved so that the indicator now points to letter “A,” correlated to the device’s detection of O in slot (a).  
\begin{figure}[ht]
\centering
\includegraphics[width=2.0 in]{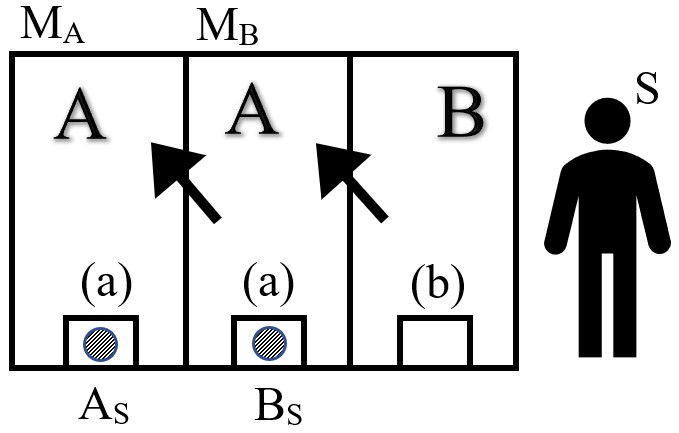}
\caption{Fig.\ 3b. By time $t_1+\Delta t$, device M evolves to a macroscopic pointer state in which an indicator points to letter “A.”  Because the device’s output is completely uncorrelated to localization at $A_S$ or $B_S$, it failed to perform the desired measurement.}
\end{figure}

However, now we have a problem.  In Fig.\ 3b, the pointer indicating “A,” which is correlated to the device’s localization of O at $A_M$, is \textit{not} correlated to the object’s localization at $A_S$.  When S reads the device’s indicator at time $t_2$, it is not that the output is certain to be wrong, but rather that the output is certain to be \textit{uncorrelated} to the measurement he intended to make.  It’s not merely that the output is unreliable – sometimes being right and sometimes being wrong – it’s that the desired measurement simply did not occur.  The correlation event at time $t_1$ did not correlate the scientist to the object’s location relative to him.  That is only possible if that location which M would measure at $t_1$ as $A_M$ could be measured by S as either $A_S$ or $B_S$, which is only possible if $A_M \neq A_S$.

To recap: At $t_0$, S sets up the experiment so that $A_M \approx A_S$ and $B_M \approx B_S$, which is what S requires so that M actually measures what it is designed to measure.  If Eq.\ 1 is correct, then at time $t_1$, $A_M \neq A_S$ and $B_M \neq B_S$.  

This has two consequences.  First, we need to explain how M could become adequately delocalized relative to S so that $A_M \neq A_S$ and $B_M \neq B_S$ in the time period from $t_0$ to $t_1$.  We are not talking about relative motion or shifts; we are talking about $A_M$ and $A_S$ becoming decorrelated from each other so as to be in a location superposition relative to each other.  In other words, how do $A_M$ and $A_S$, which were well localized relative to each other at time $t_0$, become so “fuzzy” relative to each other via quantum wave packet dispersion that $A_M \neq A_S$ at time $t_1$?  They don’t.  As long as the relative coherence length between two objects is small relative to the distance $d$ separating distinct location eigenstates, then the situation in which $A_M \neq A_S$ cannot happen over \textit{any} time period \cite{Tegmark}.  

Second, the quantum amplification in a von Neumann chain depends on the ability of measuring devices to measure what they are intended to measure.  State $\ket{A}_S + \ket{B}_S$ can only be amplified through entanglement with intermediary devices if those terms actually become correlated to states $\ket{A}_S$ and $\ket{B}_S$, respectively.  But if, as required by Eq.\ 1, there is some nonzero time period ($t_2 - t_1$) in which S can demonstrate M and O in a superposition, then the device’s measurement of the object’s location relative to it cannot correlate to the object’s location relative to S, in which case future states of S cannot be correlated to either $\ket{A}_S$ or $\ket{B}_S$.  Eq.\ 1 is internally inconsistent and is therefore false.  There is no time period in which S can measure M and O in a superposition.

This is not a subtle or merely semantic contradiction I am pointing out.  Let us place a particle in a superposition in which one eigenstate localizes the particle on the left side of a huge auditorium (in front of a giddy audience of physicists), while the other eigenstate localizes it on the right side.  I then use a measuring device – I \textit{must} use a measuring device, of course, since we are talking about a particle too tiny to see – configured so that if it measures the particle on the left side, it displays a huge “A” visible for all to see, and vice versa for the right side.  Now, if Eq.\ 1 is correct, then when we look at the device and see a huge “A” on it, \textit{it does not tell us that the particle was detected on the left side of the room!}  Rather, it tells us that the device measured it at location $A_M$ which is uncorrelated – and thus irrelevant – to our intended measurement.  If Eq.\ 1 is correct, then the device does not measure.

In conclusion, we start with the simplest and most fundamental of measurements, in which scientist S attempts to measure the position of object $\ket{O} = \ket{A}_S + \ket{B}_S$ by using measuring device M chosen so that $\ket{A}_M \approx \ket{A}_S$ and $\ket{B}_M \approx \ket{B}_S$.  We then assume that the measurement occurs by quantum amplification without collapse, as described by Eq.\ 1, in which M can entangle with O in a way that remains reversible by S for some nonzero time period.  This implies that at time $t_1$, $\ket{A}_M \neq \ket{A}_S$ and $\ket{B}_M \neq \ket{B}_S$ – i.e., the macroscopic pointer state to which M evolves is uncorrelated to the position of O relative to S.  Then at time $t_2$, the scientist in state $\ket{S_A}$ has observed device M in state $\ket{M_A}$ and the scientist in state $\ket{S_B}$ has observed device M in state $\ket{M_B}$, neither of which is correlated to the position of object O at $A_S$ or $B_S$.  Thus, Eq.\ 1 cannot yield the desired measurement.  If measurement proceeds by the quantum amplification of Eq.\ 1, then there is no measurement.

Eq.\ 1 is internally inconsistent because it implies that a measuring device used to distinguish states $\ket{A}_S$ and $\ket{B}_S$ cannot do so.  A reversible measurement is indeed a contradiction.  Therefore, if the in-principle possibility of Schr\"odinger's Cat depends on whether measurement proceeds according to quantum amplification without collapse, as characterized by Eq.\ 1, then the cat is truly dead.


\begin{thebibliography}{99}

\bibitem{Aharonov}Aharonov, Y. and Kaufherr, T., 1984. Quantum frames of reference. \textit{Physical Review D}, 30(2), p.368.
\bibitem{Aaronson}Aaronson, S., 2016. The Ghost in the Quantum Turing Machine. In: \textit{The Once and Future Turing: Computing the World.} Cambridge University Press.
\bibitem{Aaronson2}Aaronson, S., Atia, Y. and Susskind, L., 2020. On the hardness of detecting macroscopic superpositions. Preprint at https://arxiv.org/abs/2009.07450.
\bibitem{Barros}de Barros, J.A. and Oas, G., 2017. Can we falsify the consciousness-causes-collapse hypothesis in quantum mechanics?. \textit{Foundations of Physics}, 47(10), pp.1294-1308.
\bibitem{Boekholt}Boekholt, T.C.N., Portegies Zwart, S.F. and Valtonen, M., 2020. Gargantuan chaotic gravitational three-body systems and their irreversibility to the Planck length. \textit{Monthly Notices of the Royal Astronomical Society,} 493(3), pp.3932-3937.
\bibitem{Bong}Bong, K.W., Utreras-Alarc\'on, A., Ghafari, F., Liang, Y.C., Tischler, N., Cavalcanti, E.G., Pryde, G.J. and Wiseman, H.M., 2020. A strong no-go theorem on the Wigner’s friend paradox. \textit{Nature Physics}, 16(12), pp.1199-1205.
\bibitem{Brukner}Brukner, \u{C}., 2018. A no-go theorem for observer-independent facts. \textit{Entropy}, 20(5), p.350.
\bibitem{D'Ariano}D’Ariano, G.M., 2020. No purification ontology, no quantum paradoxes. \textit{Foundations of Physics}, 50(12), pp.1921-1933.
\bibitem{Deutsch}Deutsch, D., 1985. Quantum theory as a universal physical theory. \textit{International Journal of Theoretical Physics}, 24(1), pp.1-41.
\bibitem{Frauchiger}Frauchiger, D. and Renner, R., 2018. Quantum theory cannot consistently describe the use of itself. \textit{Nature communications}, 9(1), pp.1-10.
\bibitem{Gao}Gao, S., 2019. The measurement problem revisited. \textit{Synthese}, 196(1), pp.299-311.
\bibitem{GRW}Ghirardi, G.C., Rimini, A. and Weber, T., 1986. Unified dynamics for microscopic and macroscopic systems. \textit{Physical Review D}, 34(2), p.470.
\bibitem{Giacomini}Giacomini, F., Castro-Ruiz, E. and Brukner, \u{C}., 2019. Quantum mechanics and the covariance of physical laws in quantum reference frames. \textit{Nature communications}, 10(1), pp.1-13.
\bibitem{Hagar}Hagar, A., 2009. Active Fault-Tolerant Quantum Error Correction: The Curse of the Open System. \textit{Philosophy of Science,} 76(4), pp.506-535.
\bibitem{Hobson}Hobson, A., 2018. Review and suggested resolution of the problem of Schrodinger’s cat. \textit{Contemporary Physics}, 59(1), pp.16-30.
\bibitem{Joos}Joos, E., Zeh, H.D., Kiefer, C., Giulini, D.J., Kupsch, J. and Stamatescu, I.O., 2013. \textit{Decoherence and the appearance of a classical world in quantum theory.} Springer Science \& Business Media.
\bibitem{Kastner}Kastner, R.E., 2020. Unitary-only quantum theory cannot consistently describe the use of itself: on the Frauchiger–Renner paradox. \textit{Foundations of Physics}, 50(5), pp.441-456.
\bibitem{Knight}Knight, A., 2020.  No paradox in wave-particle duality.  \textit{Foundations of Physics}, 50(11), pp. 1723-27.
\bibitem{Loveridge}Loveridge, L., Busch, P. and Miyadera, T., 2017. Relativity of quantum states and observables. \textit{EPL (Europhysics Letters)}, 117(4), p.40004.
\bibitem{Maccone}Maccone, L., 2009. Quantum solution to the arrow-of-time dilemma. \textit{Physical review letters}, 103(8), p.080401.
\bibitem{Maudlin}Maudlin, T., 1995. Three measurement problems. \textit{Topoi}, 14(1), pp.7-15.
\bibitem{Neumann}von Neumann, J., 1932. Mathematical Foundations of Quantum Mechanics. Princeton University Press.
\bibitem{Penrose}Penrose, R., 1989. \textit{The Emperor's New Mind: Concerning Computers, Minds, and the Law of Physics.} Oxford University Press.
\bibitem{Proietti}Proietti, M., Pickston, A., Graffitti, F., Barrow, P., Kundys, D., Branciard, C., Ringbauer, M. and Fedrizzi, A., 2019. Experimental test of local observer independence. \textit{Science advances}, 5(9), p.eaaw9832.
\bibitem{Relano}Rela\~no, A., 2020. Decoherence framework for Wigner's-friend experiments. \textit{Physical Review A}, 101(3), p.032107.
\bibitem{Roselli}Roselli, C. and Stella, B.R., 2021. The Dead-Alive Physicist Experiment: A Case-Study Against the Hypothesis that Consciousness Causes the Wave-Function Collapse in the Quantum Mechanical Measurement Process. \textit{Foundations of Physics}, 51(1), pp.1-11.
\bibitem{Salom}Salom, I., 2018. To the rescue of Copenhagen interpretation. \textit{arXiv preprint arXiv:1809.01746.}
\bibitem{Schrodinger}Schr\"odinger, E., 1935. Die gegenwärtige situation in der quantenmechanik.  \textit{Naturwissenschaften}, 23(49), pp.823-828.
\bibitem{Tegmark}Tegmark, M., 1993. Apparent wave function collapse caused by scattering. \textit{Foundations of Physics Letters,} 6(6), pp.571-590.
\bibitem{Wigner}Wigner, E.P., 1961.  Remarks on the mind-body question.  In \textit{The Scientist Speculates} (pp.284-302). Heinemann.
\bibitem{Yin}Yin, Z.Q. and Li, T., 2017. Bringing quantum mechanics to life: from Schr\"odinger's cat to Schr\"odinger's microbe. \textit{Contemporary Physics}, 58(2), pp.119-139.
\bibitem{Zukowski}\.Zukowski, M. and Markiewicz, M., 2021. Physics and Metaphysics of Wigner’s Friends: Even Performed Premeasurements Have No Results. \textit{Physical Review Letters,} 126(13), p.130402.
\bibitem{Zych}Zych, M., Costa, F. and Ralph, T.C., 2018. Relativity of quantum superpositions. \textit{arXiv preprint arXiv:1809.04999.}

\end{thebibliography}
\end{document}